# Confidential Algorithm for Golden Cryptography Using Haar Wavelet

Marghny H. Mohamed
*Computer Science Department*
*Faculty of Computers and*
*Information*
*Assuit University*
*Egypt*

Yousef B. Mahdy
*Computer Science Department*
*Faculty of Computers and*
*Information*
*Assuit University*
*Egypt*

Wafaa Abd El-Wahed Shaban
*Information Systems Department*
*Faculty of Computers and*
*Information*
*Assuit University*
*Egypt*

*Abstract—* **One of the most important consideration techniques when one want to solve the protecting of digital signal is the golden matrix. The golden matrices can be used for creation of a new kind of cryptography called the golden cryptography. Many research papers have proved that the method is very fast and simple for technical realization and can be used for cryptographic protection of digital signals. In this paper, we introduce a technique of encryption based on combination of haar wavelet and golden matrix. These combinations carry out after compression data by adaptive Huffman code to reduce data size and remove redundant data. This process will provide multi-security services. In addition Message Authentication Code (MAC) technique can be used to provide authentication and the integrity of this scheme. The proposed scheme is accomplished through five stages, the compression data, key generation, encryption stage, the decryption stage and decompression at communication ends.**

## I. INTRODUCTION

The main challenge in data communication is focused on how to keep data secure against unlawful interference. One of the common serious attacks which threaten data security today is: intercepted; which occurs when an unauthorized party can access to read protected file and modify data. Many papers try to improve golden cryptography to solve this challenge [7, 11, 12, 13, 15]. Cryptosystems rely on the assumption that a number of mathematical problems are computationally intractable in the sense that they cannot be solved in polynomial time.

The simplicity and beauty of Fibonacci numbers have been motivated to develop matrix cryptosystems, which are useful in digital communications, i.e., digital TV, digital telephony, digital measurement, etc. One of such cryptosystems, called the golden cryptography based on the golden matrices, which are a generalization of Fibonacci Q-matrices for continuous domain, was introduced by Stakhov [13]. Any cryptosystem is considered to be secure if it is resistible against different types of cryptanalytic attacks such as the ciphertext-only attack, the known-plaintext attack and the chosen-plaintext (chosen ciphertext) attack, etc. In case of chosen plaintext attack, the cryptanalyst can obtain the ciphertexts corresponding to an arbitrary set of plaintexts of his own choosing. Unfortunately,

Rey and Sanchez [9] showed that the cryptosystem proposed [13] is not secure against chosen plaintext attack, where the secret key can be obtained easily. Another interesting cryptosystem based on Hadamard product of golden matrices was introduced by Nally [11]. There are also other simple cryptographic methods [7, 12, 15] based on extensions of golden matrices. M.Tahghighi, et al., proved that these methods are also insecure against chosen-plaintext attack [1]. So in this paper, we will try to solve the problem by proposing an improved version of golden cryptography by using Haar wavelet for golden matrix (Fibonacci Numbers, ELC Numbers and Lucas Numbers). This leads that the proposed approach own the powerful properties of the haar wavelet such as orthonormality, compact support, varying degrees of smoothness, localization both in time or space and scale (frequency), and fast implementation. In addition, one of the key advantages of wavelets is the ability to adapt to the features of a function such as discontinuities and varying frequency behavior [19, 21]. Traditional cryptographic algorithms, such as DES, AES, RSA, etc. [16, 20] send the ciphertext over the cyberspace while keeping a secret part (i.e. key) shared, which tends to be dangerous, as any intruder can get the encrypted message and apply his own cryptanalysis techniques, this means when data travel over the network even though it is hidden more attacks could be applied to the cipher message trying to get full or partial information from the message. In our scheme, the data sent over the communication channel are not the original encrypted message, but this is the compressed one and the encipher matrix generated in the sender side and the decipher matrix generated in receiver side while keeping secret part (i.e. N recurrence sequences , number of haar wavelet level and type of recurrence matrix) shared. Also, sending the HMAC of the compressed data along with the cipher enables the receiver party to verify the sender identity and message integrity. Thus, our model is carried out by several mechanisms like adaptive Huffman coding, Recurrence relations, Haar wavelet, Hash based Message Authentication Code (HMAC) to build the encryption phase.

### A. Adaptive Huffman Coding

Huffman coding needs some knowledge of the probabilities of the source sequence. If this information is unavailable, compressing the file requires two passes: the statistics are collected in the first pass, and the source is encoded in the second pass. In adaptive huffman coding convert this algorithm into a one-pass procedure, neither transmitter nor receiver knows anything about the source sequence at the start of transmission.

Both the transmitter and the receiver build tree consisting of a single node that corresponds to all symbols not yet transmitted (NYT) and has a weight of zero. As transmission progresses, nodes corresponding to symbols transmitted will be added to the tree, and the tree is reconfigured using an update procedure. Both transmitter and receiver start with the same tree structure. The updating procedure used by both transmitter and receiver is identical. Therefore, the encoding and decoding processes remain synchronized [17].

*1) Update Procedure:* The update procedure requires that the nodes be in a fixed order. This ordering is preserved by numbering the nodes. Figure 1 is a flowchart of the updating procedure [17].

*2) Encoding Procedure:* The flowchart for the encoding procedure is shown in Figure 2 [17].

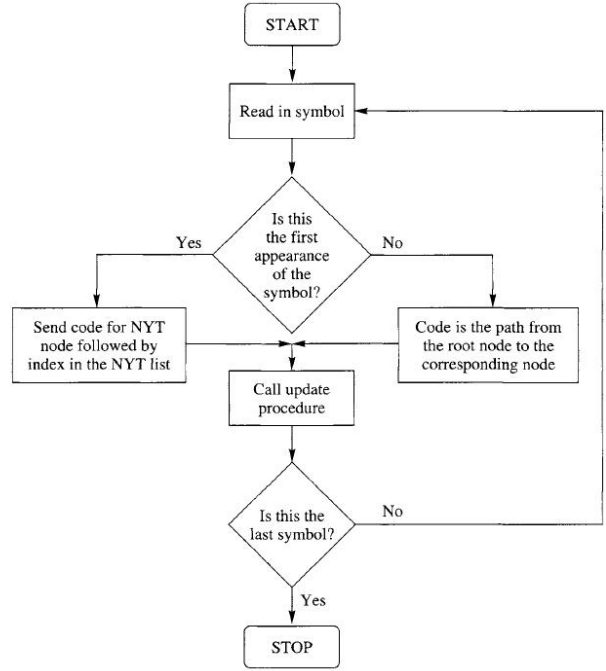

Figure 2. Flowchart off the encoding procedure

## B. RECURRENCE RELATIONS

Recurrence relation is useful in certain counting problems like Fibonacci numbers, Lucas and ELC. A recurrence relation relates the nth element of a sequence to its predecessors. Recurrence relations are related to recursive algorithms. A recursive relation for the sequence $a_0; a_1; a_2; \ldots\ldots$ is an equation that relates $a_n$ to certain of its preceding terms $a_0; a_1; a_2; \ldots.. a_{n-1}$.

Initial conditions for the sequence $a_0; a_1; a_2\ldots..$ are explicitly given values for a finite number of the terms of the sequence.

In this section recurrence relations Fibonacci, Lucas and ELC numbers were presented and their application to cryptography is examined [7, 12-13, 22-24].

*1) Fibonacci numbers:* Fibonacci numbers are given by the following recurrence relation [22-23]

$$F_{n+1} = F_n + F_{n-1} \qquad (1)$$

With the initial conditions

$$F_1 = F_2 = 1 \qquad (2)$$

A square matrix ($2 \times 2$) as shown below was introduced in [22]

$$Q = \begin{pmatrix} 1 & 1 \\ 1 & 0 \end{pmatrix} \qquad (3)$$

The following property of the nth power of the Q-matrix was proved

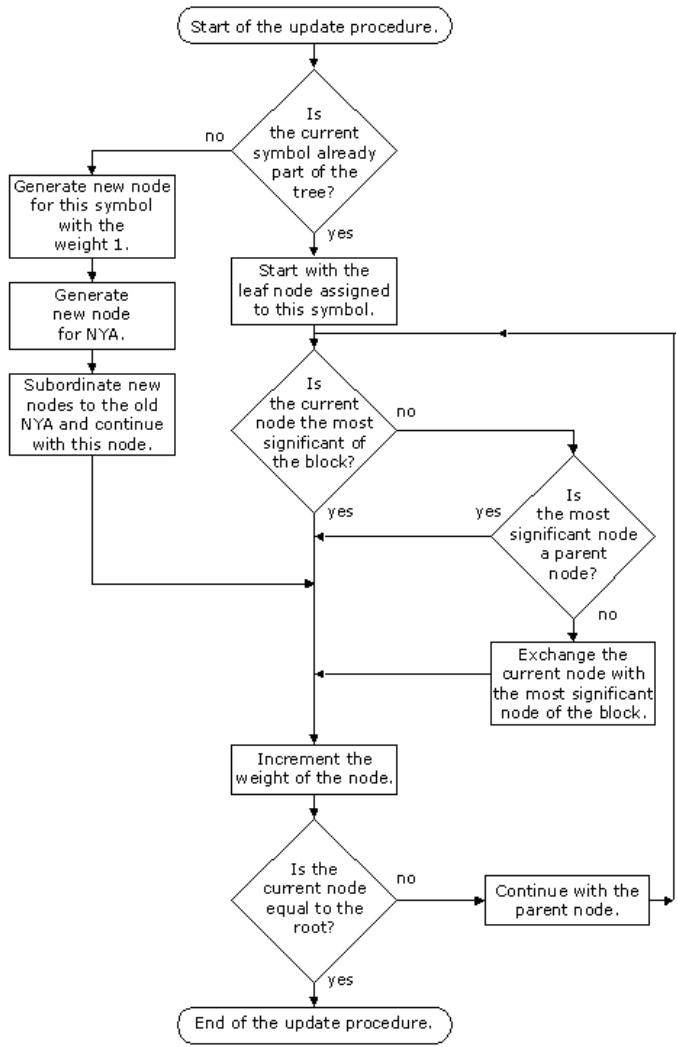

Figure 1. Update procedure for the adaptive Huffman coding algorithm.

$$Q^n = \begin{pmatrix} F_{n+1} & F_n \\ F_n & F_{n-1} \end{pmatrix} \qquad (4)$$

Where n = 0, ±1; ±2; ±3; . . ., $F_{n-1}$; $F_n$; $F_{n+1}$ are Fibonacci numbers.

Hence, the inverse of matrices Qn is

$$Q^{-n} = \begin{pmatrix} +F_{n-1}/(-1)^n & -F_n/(-1)^n \\ -F_n/(-1)^n & +F_{n+1}/(-1)^n \end{pmatrix} \qquad (5)$$

The generalized Fibonacci matrix Qp is defined by

$$Q_p = \begin{pmatrix} 1 & 1 & 0 & . & . & . & . & . & 0 \\ 0 & 0 & 1 & . & . & . & . & . & 0 \\ . & . & . & . & . & . & . & . & . \\ . & . & . & . & . & . & . & . & . \\ . & . & . & . & . & . & . & . & . \\ . & . & . & . & . & . & . & . & . \\ . & . & . & . & . & . & . & . & . \\ 0 & 0 & 0 & . & . & . & . & 0 & 1 \\ 1 & 0 & 0 & . & . & . & 0 & 0 & 0 \end{pmatrix} \qquad (6)$$

Note that the Qp-matrix is a square (p + 1) × (p + 1) matrix.

For p = 0, 1, 2, 3 ... the Qp-matrices have the following forms, respectively:

$$Q_0 = \begin{pmatrix} 1 \end{pmatrix} \qquad Q_1 = \begin{pmatrix} 1 & 1 \\ 1 & 0 \end{pmatrix} \qquad (7)$$

$$Q_2 = \begin{pmatrix} 1 & 1 & 0 \\ 0 & 0 & 1 \\ 1 & 0 & 0 \end{pmatrix} \qquad Q_3 = \begin{pmatrix} 1 & 1 & 0 & 0 \\ 0 & 0 & 1 & 0 \\ 0 & 0 & 0 & 1 \\ 1 & 0 & 0 & 0 \end{pmatrix} \qquad (8)$$

In general the nth power of the Qn matrix

$$Q_p^n = \begin{pmatrix} F_p(n+1) & F_p(n) & \cdots & F_p(n-p+2) & F_p(n-p+1) \\ F_p(n-p+1) & F_p(n-p) & \cdots & F_p(n-2p+2) & F_p(n-2p+1) \\ \vdots & \vdots & \ddots & \vdots & \vdots \\ F_p(n-1) & F_p(n-2) & \cdots & F_p(n-p) & F_p(n-p-1) \\ F_p(n) & F_p(n-1) & \cdots & F_p(n-p+1) & F_p(n-p) \end{pmatrix}$$

*2) Lucas numbers:* The sequence of Lucas numbers $L_k$ is defined by the second-order linear recurrence formula and initial terms

$$L_{k+1} = L_k + L_{k-1}, L_0 = 2, L_1 = 1 \qquad (9)$$

The proposed matrix using Lucas recursion

$$L^n = \begin{pmatrix} L_{n+1} & L_n \\ L_n & L_{n-1} \end{pmatrix} \qquad (10)$$

$$L^1 = \begin{pmatrix} 2 & 1 \\ 1 & 3 \end{pmatrix} \qquad (11)$$

The inverse of matrices $L^n$ is

$$L^{-n} = \begin{pmatrix} +L_{n-1}/((-1)^n * 4) & -L_n/((-1)^n * 4) \\ -L_n/((-1)^n * 4) & +L_{n+1}/((-1)^n * 4) \end{pmatrix} \qquad (12)$$

The other explicit forms of Ln can be obtained recursively same as $Q^n$

*2) ELC numbers:* ELC numbers are given by the following recurrence relation $E_{n+1} = E_n + E_{n-1}$, with condition $E_0 = 8$ and $E_1 = 14$. The golden matrix using ELC recursion is proposed as follows.

$$E^n = \begin{pmatrix} E_{n+1} & E_n \\ E_n & E_{n-1} \end{pmatrix} \qquad (13)$$

Where n = 0, 1, 2, 3, the inverse of matrices $E^n$ is

$$E^{-n} = \begin{pmatrix} +E_{n-1}/((-1)^n * 20) & -E_n/((-1)^n * 20) \\ -E_n/((-1)^n * 20) & +E_{n+1}/((-1)^n * 20) \end{pmatrix} \qquad (14)$$

The other explicit forms of $E^n$ can be obtained recursively same as $Q^n$

## C. Haar Wavelet Transform

Haar wavelet is the simplest wavelet. Haar transform or Haar wavelet transform has been used as an earliest example for orthonormal wavelet transform with compact support [19, 21]. The Haar wavelet transform is the first known wavelet and was proposed in 1909 by Alfred Haar. The Haar wavelet transform has a number of advantages:

- It is conceptually simple.

- It is fast.

- It is memory efficient, since it can be calculated in place without a temporary Array.

*1) Procedure for Haar Wavelet Transform:* To calculate the Haar transform of an array of n samples [18]:

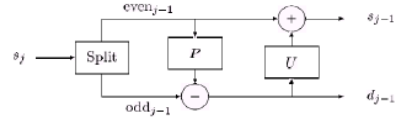

1) *Split*: divide the input data into:
- Even indexed samples $S_n$.
- Odd indexed samples $S_{n+1}$. Lazy wavelet transform

2) Predict: the odd elements from the even elements-output detail.

$$d_{j-1} = s_j[2n+1] - s_j[2n]. \qquad (15)$$

In general:

$$d_{j-1} = odd_{j-1} - P(even_{j-1}). \qquad (16)$$

3) Update:
- Follows the predict phase.
- The approximations $S_{n-1}$(the signal for next step) should maintain the average of the original signal $S_n$.

$$s_{j-1}[n] = s_j[2n] + d_{j-1}[n]/2. \qquad (17)$$

In general:

$$s_{j-1} = even_{j-1} + U(d_{j-1}). \qquad (18)$$

## D. Hash-*based* Message Authentication Code (HMAC)

The purpose of an MAC is to authenticate both the source of a message and its integrity without the use of any additional mechanisms. HMACs have two functionally distinct parameters, a message input and a secret key known only to the message originator and intended receiver(s). Additional applications of keyed-hash functions include their use in challenge-response identification protocols for computing responses, which are a function of both a secret key and a challenge message. An HMAC is used by the message sender to produce a value (the MAC) that is formed by condensing the secret key and the message input. The MAC is typically sent to the message receiver along with the message [8]. The receiver computes the MAC on the received message using the same key and HMAC function as was used by the sender, and compares the result computed with the received MAC. If the two values match, the message has been correctly received, and the receiver assures us that the sender is a member of the community of users that share the key, as shown in Figure 3.

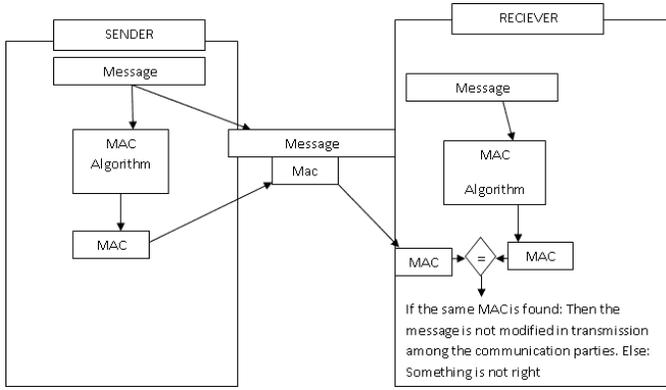

Figure 3.    The use of MAC

HMAC-SHA-256+ are secret key algorithms. While no fixed key length is specified in [HMAC], key lengths less than the output length decrease security strength, and keys longer than the output length do not significantly increase security strength [14].

The rest of this paper is organized as following, the proposed scheme is presented in section 2, in section 3 the security analysis is introduced, experimental results is presented in section 4, finally conclusions are provided in section 5.

## II.    THE PROPOSED SCHEME

This section examines improving golden cryptography to solve insecurity against the chosen-plaintext attack for golden cryptography by using haar wavelet transform, in terms of the problems the secret key can be obtained for example, A.Stakhov suggested"new kind of cryptography system" in [13] but Rey and Sanchez showed that this cryptosystem is not secure against chosen plaintext attack [9], which they let pairs of plaintext M1, M2, M3 and M4, which

$$M1 = \begin{pmatrix} 1 & 0 \\ 0 & 0 \end{pmatrix}, M2 = \begin{pmatrix} 0 & 1 \\ 0 & 0 \end{pmatrix}$$
$$, M3 = \begin{pmatrix} 0 & 0 \\ 1 & 0 \end{pmatrix}, M4 = \begin{pmatrix} 0 & 0 \\ 0 & 1 \end{pmatrix}$$

And the enciphering matrix is $Q^{2x}$

$$Q^{2x} = \begin{pmatrix} cFs(2x+1) & sFs(2x) \\ sFs(2x) & cFs(2x1) \end{pmatrix}$$

For a more detailed description of such functions we refer the reader to [1, 9]. Using simple calculus shows, the real value for x is:

$$x = \frac{1}{2}\log_\tau\left(\frac{\sqrt{k_1\sqrt{5} + \sqrt{5k_1^2 + 4}}}{2}\right), \quad (19)$$

where $k_1 \in R$ can be obtained from the equation

$$z = \frac{k_1\sqrt{5} \pm \sqrt{5k_1^2 + 4}}{2}, \quad (20)$$

Where $z = \tau^{2t}$. $\tau$ is golden proportion, thus the secret key x is obtained [9]. By the similar calculations M.Tahghighi, et al. proved that these methods [7, 12, and 15] insecure against chosenplaintext attack [1], which the secret key x of the Hadamard product of golden matrices can be obtained by:

$$x = \frac{1}{4}\log_\tau\left(\frac{2 - 2.5(t_1 - t_2) \pm \sqrt{6.25(t_1 - t_2)^2 + 5(t_1 - t_2)}}{2}\right), \quad (21)$$

known real variables $t_1$ and $t_2$, we obtain the system

$$Fs(2x+1)cF(2x1) - [sFs(2x)]^2 = t_1, \quad (22)$$

$$cFs(2x+1)cF(2x-1) + [sFs(2x)]^2 = t_2, \quad (23)$$

of non-linear equations. By the definition of $sF(x)$ and $cFs(x)$,

$$\frac{(\tau^{2t+1} + \tau^{-2t-1})}{5} - \frac{(\tau^{2t} - \tau^{-2t})}{5} = t_1, \quad (24)$$

$$\frac{(\tau^{2t+1} + \tau^{-2t-1})}{5} + \frac{(\tau^{2t} - \tau^{-2t})}{5} = t_2, \quad (25)$$

and the secret key $\{k, x\}$ is obtained of the generalized golden cryptographic method it is easy to see that

$$k = \frac{\sigma_k^2 - 1}{\sigma_k}, \quad (26)$$

we can calculate the value of x, i.e.,

$$x = \frac{1}{2}\log\sigma_k(t_1\sigma_k + t3) \ (or \ x = \frac{1}{2}\log\sigma_k(t_1\sigma_k^{-1} + t_2)), \quad (27)$$

they are very simple and it is very easy to show their insecurity against the chosen-plaintext attack [1].

The proposed scheme can be summarized in the following stages: At *sender side* some stages must be done:

- Compression data

*1)  Map each character in plaintext into its corresponding ASCII code, M'=ASCII(M).*

*2)  Compression M' by using adaptive Huffman Coding and generate compressed data CM.*

- Producing MAC Message

*3)  Generate key by using any encryption algorithm (e.g DES, TripleDES, AES) to compute a MAC over the compressed data message CM using the HMAC function.*

- Encryption Stage

*1)  Input a cryptographic key, K=n, r.*

*2)  Construct the corresponding "Golden matrix" G Matrix Depending on r equivalent $Q^n{}_p$ or $L^n{}_p$ or $E^n{}_p$ (calculate p depended by message size p=ceil($\sqrt{M}$).*

*3)  Compute key encryption matrix E, where equivalent Haar wavelet matrix from G Matrix according to l this create random and add another number matrix.*

*4)  Break up CM into CG groups (each group contains rowmatrix2 elements) and from each on a square matrix.*

*5)  For each group, compute the corresponding CipherText where Ci = CG*E.*

*6)  Collect Ci and send its.*

Algorithm.1 shows the steps at sender side.

```
Data: M, n, l, r
where M is Plaintext, n is used as short session key(one
    time pad) and l is level, r type of Recurrence matrix
Result: chiphertext
initialization;
/* Map each character in plaintext into
    its corresponding ASCII code.        */
M' ←— ASCII(M);
/* Compression M' by using adaptive
    Huffman Coding and generate
    compressed data CM .                 */
CM ←— adaptiveHuffmanCoding(M');
/* Producing MAC Message :Generate key
    by using any encryption algorithm
    (e.g DES; TripleDES, AES) to compute
    a MAC over the compressed data
    message CM using the HMAC function.
    */
ComputeHash(CM);
/* Generate Encryption Key           */
/* Golden matrix G Matrix Depending on
    r equivalent Qₚⁿ or Lₚⁿ or Eₚⁿ       */
G ←— GeneralGoldenMatrix(n, p);
/* Compute key encryption matrix E,
    where equivalent Haar wavelet matrix
    from G Matrix according to l and add
    another number matrix.               */
E ←—
    WaveletTransform(G, l) + anothernumbermatrix;
while not at end of CM do
    if
        (CM.Length − CM.Position) > (E.Row ∗ E.Col)
        then
          │  bytesToRead = (E.Row ∗ E.Col);
        else
          │  bytesToRead = (CM.Length − CM.Position);
    CM.Read(buf, 0, bytesToRead);
    while not at end of buf do
        index=0 ;
        for i ←— 0 to E.Row do
            for j ←— 0 to E.Col do
                if index >= buf.Length then
                  │  m2[i, j] ←— -1;
                else
                  │  m2[i, j] ←— buf[index];
                  │  index=index + 1;
                end
            end
        end
        m3 ←— Matrix.Multiply(m2, E);
    end
    /* Collect m3 in chiphertext.         */
    chiphertext ←— m3;
end
```

**Algorithm 1**: Algorithm steps at sender side

```
Data: CM
where CM is Chiphertext
Result: PlainText
initialization;
/* Generate Decryption Key              */
/* Compute (IE) inverse of E matrix
    where equivalent Haar wavelet matrix
    from G Matrix.                       */
IE ←— InverseE();
while not at end of CM do
    Cipher ←— CM.read();
    while not at end of Cipher do
        index=0 ;
        for i ←— 0 to IE.Row do
            for j ←— 0 to IE.Col do
                m2[i, j] ←— Cipher[index];
                index=index + 1 ;
            end
        end
        m3 ←— Matrix.Multiply(m2, IE);
    end
    /* Collect m3 in Plaintext.           */
    CPlaintext ←— m3;
end
/* DeCompression M' by using adaptive
    Huffman Coding and generate
    compressed data CM .                 */
Plaintext ←—
    adaptiveHuffmanCodingDecode(CPlaintext);
/* Compare the obtained MAC value with
    the MAC value of the constructed
    message Obt-MAC, if the matching
    obtained (MAC = Obt-MAC). This
    indicates that the message is not
    modified in transmission among the
    communication parties.               */
Obt − MAC ←— ComputeHash(CPlaintext);
if Obt-MAC.SequenceEqual(MAC) then
  │  "Equality" ;
else
  │  "Data attack" ;
end
```

**Algorithm 2**: Algorithm steps at receiver side

### III. EXPERIMENTAL RESULTS

In order to evaluate the effectiveness of the proposed scheme, the following experiment has been conducted to measure the level of confusion and diffusion, by comparing plain to cipher the relationship as a metric model for security. These simulation experiments have been done on a sentence M representing the original message: M ="Cryptographist is the science of overt secret writing", to encrypt this message by the proposed model. Suppose k= 5, 2. The contrast between plaintext and ciphertext is demonstrated in Figure.4.

At the *receiver side* another steps are done to decrypt the ciphertext and retrieve the original message, in addition to ensuring authenticity and message integrity according to the following stages:

- Decryption Stage

1) *Compute inverse of E matrix.*

2) *Break up the ciphertext into CG groups and a square matrix for each block.*

3) *For each group, compute CG =Ci ∗E−1.*

4) *Collect CG, where Compressed message.*

5) *Decompress CG to Original message M'.*

6) *Map each ASCII in M' to corresponding character M.*

- Verification Stage

1) *Compute MAC value of the obtained Compressed message (Obt-MAC).*

2) *Compare the obtained MAC value with the MAC value of the constructed message Obt-MAC, if the matching obtained (MAC = Obt-MAC). This indicates that the message is not modified in transmission among the communication parties.*

*Algorithm.2, shows the steps at receiver side.*

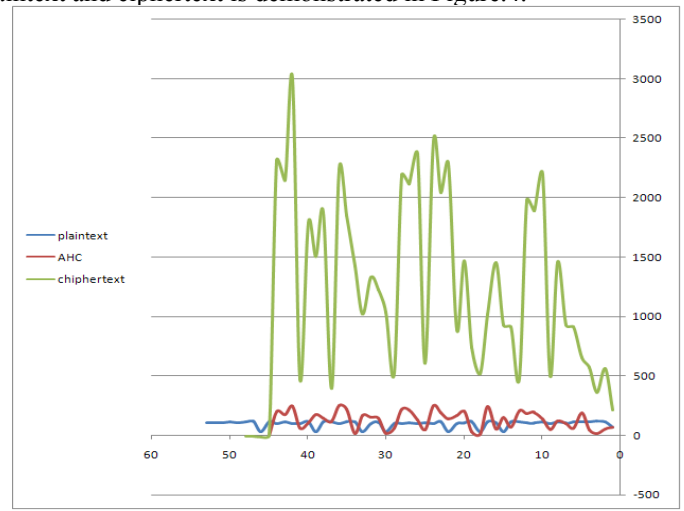

Figure 4. The Contrast between Plaintext and Ciphertext

From the resulted ciphertext we can clearly notice that for each character on the original message there is a different value appeared in the ciphertext, and there is no direct relationship between the plaintext and the cipher text. The benefits of use adaptive huffman code are reducing data and removing redundant data. This indicates that the proposed model has a high confusion because the relationship between the input (key) and the output (message) is nonlinear. We observed that the message has some repeated characters such as character"e" for example (repeated six times), and every time the resulted cipher is different from the other, the repeated values disappeared on the resulted ciphertext. Figure .5 shows the distribution of the"e" character in the ciphertext. This indicates that the proposed model provides a high-level of diffusion.

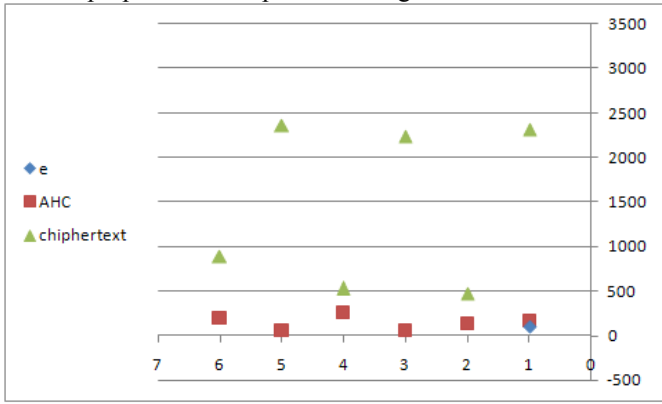

Figure 5.    Distribution of character 'e' on ciphertext

A similar experiment has also been conducted to a sentence consisting of consecutive m's as a plaintext with the length = 10 on this message M1 = "mmmnmmmomm". As we can see clearly, the resulted ciphertext is completely different from the ciphertext although the M1 character is eight times in a sentence. The contrast between plaintext and ciphertext is demonstrated in Figure.6.

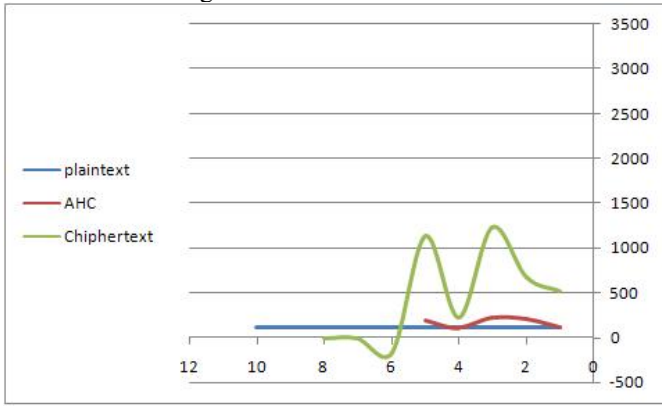

Figure 6.    The Contrast between Plaintext and Ciphertext

To confirm our results one more experiment is conducted. We encrypted another message similar to the previous one using the same key used before, to see what happens when two very similar texts are encrypted under the same key. These simulation experiments have been done on a sentence M2 representing the original message: M2 ="meet me after party meet me after party", again we can see, the resulted cipher is totally different from the previous experiment as shown in Figure.7, although it is the same message and encrypted under similar key.

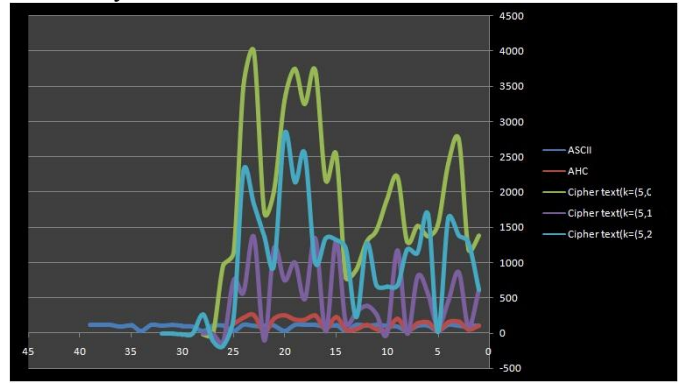

Figure 7.    The Contrast between the same message and its Ciphertext under similar keys

In order to evaluate the effectiveness of modify our proposed scheme to increase security and increase confusion, the following experiment has been conducted to measure the level of confusion and diffusion, by comparing plain to cipher relationship as a metric model for security about simulation experiments have been done on a sentence M2 representing the original message: M2 ="meet me after party meet me after party". This experiment has been conducted to measure the level of confusion in modifying our scheme by generating matrix key different size depending on level random generate and p about $Q_p^n$ or $L_p^n$ or $E_p^n$, this show in Figure .8.

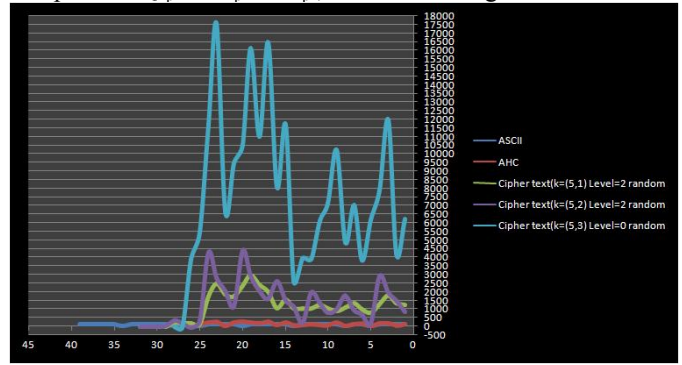

Figure 8.    The Contrast between Plaintext and Ciphertexts

In Figure.9 Comparison between schema and modify schema, we notice increase the level of confusion from the previous scheme between CipherTexts

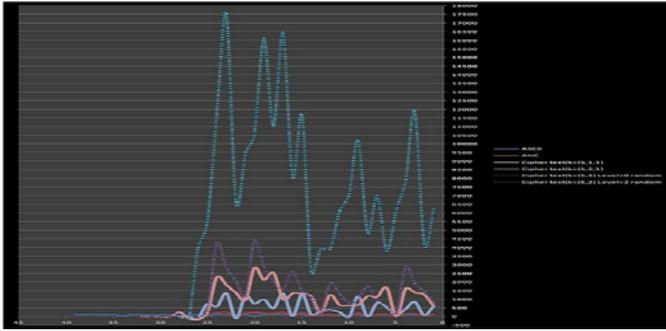

Figure 9.    Comparison between schema and modify schema

Even under this schema and modify schema, no relation between plaintext and ciphertext can be noticed, and the distribution of ciphertext is random. When we increase the message length with repetition, no relation could be noticed between ciphertext and plaintext indicates the strength of this scheme against partially known plaintext attack. This confirms what we mentioned the confusion and diffusion properties are provided by the proposed scheme. The performance of the secret key algorithms has been compared on different data, by using input file data of varying sizes and formats.

## IV.    SECURITY ANALYSIS

Some security analysis has been performed on the proposed encryption scheme [2-6 and 10], such as:

- *Known-plaintext attack*: Suppose the intruder knows some pairs of ciphertexts and corresponding plaintexts, here his goal is to reveal the shared data (keys), to use it in future to decipher other ciphertext. The intruder will then have to search in a semi-impossible search space. Consider the Haar wavelet in the proposed scheme; the attacker must firstly construct the set of possible key space. If we suppose that the key size of the Haar wavelet is $Z*Z$, where $Z$ is equal $2^{level}$ if the number of golden row matrix is already a $2^{level}$ otherwise $Z$ is the next bigger number that's a $2^{level}$, where level and p is depending on the data size used. An example to illustrate the enciphering matrix: Let K=1, 1 where N = 1, R = 1, P = 0, L = 1

1) Generate recursion matrix
   $Q_0^1 = \begin{pmatrix} 1 \end{pmatrix}$ .
2) Add padding to original matrix and calculate extra row, col
   $E = \begin{pmatrix} 1 & 0 \\ 0 & 0 \end{pmatrix}$ .
3) Row transform
   $E = \begin{pmatrix} 1 & -1 \\ 0 & 0 \end{pmatrix} \rightarrow E = \begin{pmatrix} 0.5 & -1 \\ 0 & 0 \end{pmatrix}$
4) Column transform
   $E = \begin{pmatrix} 0.5 & -1 \\ -0.5 & 0 \end{pmatrix} \rightarrow E = \begin{pmatrix} 0.25 & -1 \\ -0.5 & 0 \end{pmatrix}$
   $\rightarrow E = \begin{pmatrix} 0.25 & -1 \\ -0.5 & 1 \end{pmatrix}$ .
5) Cryptography key
   $E = \begin{pmatrix} 0.25 & -0.5 \\ -0.5 & 1 \end{pmatrix}$ .
When change L to 2
   $E = \begin{pmatrix} 0.0625 & -0.1250 & -0.5000 & 0.0000 \\ -0.1250 & 0.2500 & 0.0000 & 0.0000 \\ -0.5000 & 0.0000 & 1.0000 & 0.0000 \\ 0.0000 & 0.0000 & 0.0000 & 0.0000 \end{pmatrix}$ .

To apply scheme on this matrix, we can get more than enciphering matrices using the same key by adding random matrices to covering the 0's in E, and this example shows that the enciphering matrices size increase with an increasing level of haar wavelet, where this is exponential increasing. In addition to the one way property of her wavelet is nonlinear makes computing the key of it (inputs) difficult enough even if the intruder knows the ciphertext and the plaintext, which is not the original text, but this is the compressed one.

- *Ciphertext-only attack:* Suppose the intruder can eavesdrop the ciphertext in transmit, his goal here is to reveal the keys (inputs), or the corresponding plaintext (outputs), then the intruder must search in the key space of inputs, and in the key space of inputs is $(z*z)$ this is two dimensions matrix depending on the level of wavelet transformation, p of recurrence matrix and type of recurrence use (Fibonacci, ELC, Lucas), then the intruder must search in the following key space: $(z*z)$ and random matrix, the bigger Z and random matrix are the larger the key space which search on it. Also the plaintext is compressed data the time complexity of an adaptive Huffman encoding is linear: $N\lceil \Sigma \rceil + log(2\Sigma - 1)\rceil + Sn$ where N is the total number of input symbols, $\sum$ is the current number of unique symbols, and S is the time required, if necessary, to rebalance the tree [26, 27], and encryption of compressed data each block consists of the $T_e = G(Z^3 \triangle t_m + Z^2(Z-1)\triangle t_a)$ where $\triangle t_m$ is a time of one multiplication and $\triangle t_a$ is a time of one addition. The data complexity about encryption is O(N log$|\Sigma|$)+O($2^{level}$)+O($Z^3$) Due to this fact the cipher difficult to be broken.

- *Confusion and diffusion:* Confusion and diffusion are two basic design criteria for encryption algorithms [25]. Diffusion means spreading out the influence of a single plaintext symbol over many ciphertext symbols so as to hide the statistical structure of the plaintext. Confusion means the use of transformations to complicate the dependence of the statistics of ciphertext on that of the plaintext. The proposed cryptosystem has a high confusion and diffusion properties, which makes the cryptosystem of high key sensitivity and plaintext sensitivity, and this of high computing security. On the other hand, the mapping function of the used Golden cryptography is a nonlinear function which makes the relationship between the plaintext, key and ciphertext nonlinear. This property complicates possibility of retrieving one of them even if the others were known [8].

- *Statistical analysis:* Correlation Coefficient Analysis and t-test Statistical analysis such as correlation coefficient factor is used to measure the relationship between two variables. This factor examines the proposed encryption algorithm which strongly resists statistical attacks. Therefore, ciphertext must be completely different from the plaintext. The paired t-test and correlation use the same type of data; it is easy to confuse the two techniques. The paired t-test is used to test for differences in the mean values of each variable, while correlation shows associations between the pairs of values. If the correlation coefficient equals one, that means the plaintext and its encryption is identical. If the correlation coefficient equals zero, that means the ciphertext is completely different from the plaintext (i.e. good encryption). If the correlation coefficient equals minus one that means the ciphertext is the negative of the plaintext. So, success of the encryption process means smaller values of the correlation coefficient. The experimental results, the correlation coefficient value and Paired t-test of the proposed encryption algorithm is:

TABLE I.    THE CORRELATION AND PAIRED T-TEST

| The Correlation and paired t-test from encrypted msg2 | | |
|---|---|---|
| *meet me after the party meet me after the party* | | |
| *Ciphertext* | *Correlation* | *Paired t-test* |
| C1m1 | 0.401154711 | 0.0001 |
| C1m2 | 0.24715579 | 0.0001 |
| C2m1 | 0.401826534 | 0.0001 |
| C2m2 | 0.286313711 | 0.0001 |
| C3m1 | 0.376013828 | 0.0001 |
| C3m2 | 0.152848002 | 0.0001 |
| Rijndael | 0.260303958 | 0.08258 |
| DES | 0.446170523 | 0.0040 |
| TripleDES | -0.400377119 | 0.0775 |

TABLE II.    THE UNPAIRED T-TEST

| The Unpaired t-test from encrypted msg2 | | |
|---|---|---|
| *meet me after the party meet me after the party* | | |
| | *C1m2* | *C2m2* | *C3m2* |
| *C1m1* | 0.0006 | 0.0001 | 0.0002 |
| *C2m1* | 0.0006 | 0.0001 | 0.0002 |
| *C3m1* | 0.0007 | 0.0001 | 0.0002 |

Where C1m1, C1m2, C2m1, C2m2, C3m1 and C3m2 are chiphertexts under the same key and different random matrix. In case of chiphertext the values of correlation coefficient between 0.152848002 and 0.401154711, which means that the schema is uncorrelated. The values of t-test show that this difference is considered an extremely statistically significant; this means that the proposed encryption algorithm has a strong security.

## V.    CONCLUSION

This paper is proposing a cryptosystem based on hybrid approaches proposed. The proposed cryptosystem provided multi-security services such as confidentiality, authentication, and integrity, which there are important security services in most applications. To serve data confidentiality by using a technique of encryption based on combination of haar wavelet and golden matrix. These combinations are carried out after compression data by adaptive Huffman code to reduce data size, remove redundant data and consider this initial encryption to the data because it becomes more sensitive to transmission errors or any change in data by an intruder. MAC technique produced the digital signature of the scheme for providing the other mentioned security services. In this scheme the digital signature firstly produced by computing the MAC of the compressed message, then signing it by the sender's private key generated by any encryption algorithm. Finally, the signed MAC and cipher will be sending to cyberspace. At the receiver end, after the decryption is done, the digital signature about compressed data can be used to verify the integrity of the message and the authentication of the sender then decompress the data to get at the original data. The experimental results indicate that the proposed cryptosystem has a high confusion and diffusion properties, it has high security and it is suitable for secure communications.